\documentclass[journal,11pt,onecolumn,draftclsnofoot,]{IEEEtran}
\IEEEoverridecommandlockouts

\usepackage{cite}
\usepackage{amsmath,amssymb,amsfonts}
\usepackage{algorithmic}
\usepackage{graphicx}
\usepackage{textcomp}
\usepackage{xcolor}
\usepackage{import}
\usepackage{tabularx}
\usepackage{array}
\usepackage{tabu}
\usepackage{multirow}
\usepackage{siunitx}
\usepackage{booktabs}

\usepackage{fullpage}
\usepackage{times}
\usepackage{fancyhdr,graphicx,amsmath,amssymb}
\usepackage[ruled,vlined]{algorithm2e}
\include{pythonlisting}

\def\BibTeX{{\rm B\kern-.05em{\sc i\kern-.025em b}\kern-.08em
    T\kern-.1667em\lower.7ex\hbox{E}\kern-.125emX}}

\newcommand\Topspace{\rule{0pt}{4ex}}           
\newcommand\Bottomspace{\rule[-2ex]{0pt}{0pt}}  

\renewcommand{\baselinestretch}{0.94}
\begin{document}

\title{A Solution for Dynamic Spectrum Management in Mission-Critical UAV Networks
}


\author{
	\IEEEauthorblockN{Alireza Shamsoshoara\IEEEauthorrefmark{1}, Mehrdad Khaledi \IEEEauthorrefmark{2}, Fatemeh Afghah \IEEEauthorrefmark{1}, Abolfazl Razi \IEEEauthorrefmark{1},\\ Jonathan Ashdown\IEEEauthorrefmark{3}, Kurt Turck \IEEEauthorrefmark{3}
	}
	\\
	\IEEEauthorblockA{\IEEEauthorrefmark{1}School of Informatics, Computing and Cyber Systems, Northern Arizona University, Flagstaff, AZ, USA}
	
	\IEEEauthorblockA{\IEEEauthorrefmark{2} Mathematics \& Computer Science Department, Suffolk University, Boston, MA, USA}
	\IEEEauthorblockA{\IEEEauthorrefmark{3} Air Force Research Laboratory, Rome, NY, USA} }

\maketitle
\begin{abstract}

In this paper, we study the problem of spectrum scarcity in a network of unmanned aerial vehicles (UAVs) during mission-critical applications such as disaster monitoring and public safety missions, where the pre-allocated spectrum is not sufficient 
to offer a high data transmission rate for real-time video-streaming. In such scenarios, the UAV network can lease 
part of the spectrum of a terrestrial licensed network in exchange for providing relaying service. In order to optimize the performance of the UAV network and prolong its lifetime, some of the UAVs will function as a relay for the primary network while the rest of the UAVs carry out their sensing tasks. Here, we propose a team reinforcement learning algorithm performed by the UAV's controller unit to determine the optimum allocation of sensing and relaying tasks among the UAVs as well as their relocation strategy at each time. We analyze the convergence of our algorithm and present simulation results to evaluate the system throughput in different scenarios.

\end{abstract}


\section{Introduction}\label{Introduction}
Unmanned Aerial Vehicles (UAV) have several advantages such as flexibility, agility, wide field of imaging view and low deployment cost over ground-based infrastructures; hence, are widely used in many applications such as surveillance, disaster relief, wildlife monitoring, and military applications \cite{Gartner_UAV,Khaledi_SECON18,mousavi2018leader,mousavi2019use,peng2018unified,rovira2019optimized}. 
Also, UAVs face many security challenges based on the vast area of applications. Many methods such as key sharing, authentication, authorization, and so on have been proposed for these devices in the past few years \cite{shamsoshoara2019overview,shamsoshoara2019ring}.
In some of these missions such as disaster monitoring and military applications, the pre-allocated spectrum for UAVs is not sufficient to collect and transmit high-throughput imagery data in certain locations or times, hence an on-demand access to additional spectrum is required. In this paper, we propose a solution for securing extra spectrum for the UAV networks.

In this solution, the UAV network leases a portion of time access to the licensed spectrum of a terrestrial network. In order to compensate the spectrum access grant, some of the UAVs provide relaying service for the terrestrial network, while the rest of UAVs perform their regular sensing and transmission tasks using the leased spectrum. The allocation of relaying and sensing tasks among the UAVs as well as selecting their motion strategy is performed by a high-capability drone which serves as controller unit of the entire network. This optimization is performed considering various factors such as the position of the UAVs and ground0based nodes, the quality of communication channels as well as the expected lifetime of the drones.

Cooperative spectrum leasing as a property-right model spectrum sharing solution has been studied previously in the context of cognitive radio networks \cite{Stanojev08,Afghah_CDC2013,Namvar15,valehi2017maximizing,valehi2018online,razi2017delay}. The main objective of these works is maximizing the benefits of the primary network through optimal time allocation between the primary and secondary networks. The ignorance of the performance of the secondary network, can cause significant operation problems in critical missions.

\begin{figure*}[hbt!]
\centering
\includegraphics[width=0.72\paperwidth]{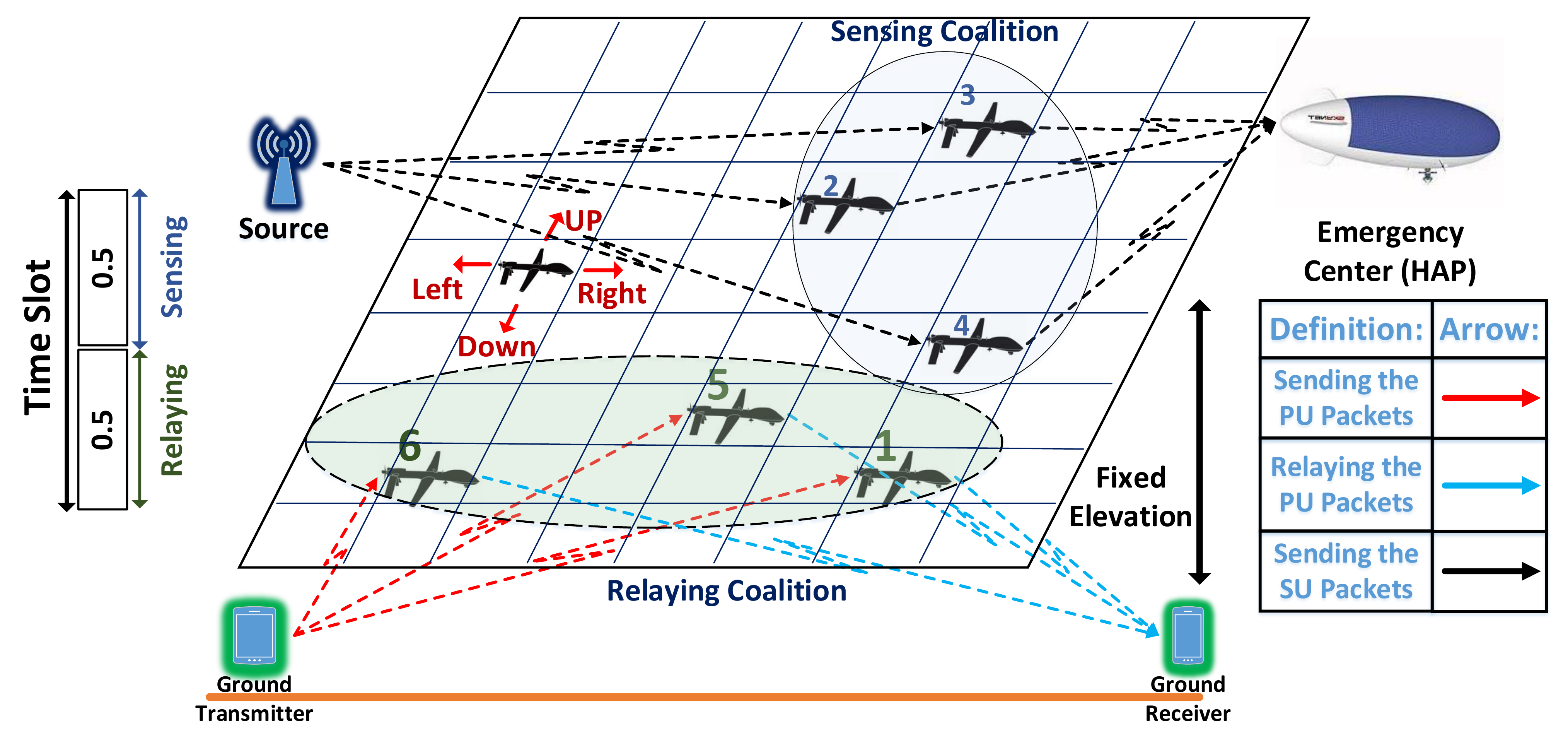}
\caption{A sample scheme of the proposed spectrum sharing between a UAV network and a terrestrial licensed network.}
\label{fig:system_grid_1}
\end{figure*}

The task allocation problem to select UAVs for relaying or sensing tasks can be modeled as a multi-agent reinforcement learning (MARL) problem \cite{Tan93}. In a cooperative MARL problem, multiple learning agents work together to achieve a system-wide optimal solution. There have been efforts to solve task coordination in a fully distributed manner. However, the proposed solutions impose a heavy computation and decision making loads to the UAVs, and still may not converge to an optimal solution. Hence, distributed task allocation solutions are not practically feasible in highly dynamic systems \cite{kok2006collaborative,shamsoshoara2019distributed}. 
In this paper, we consider a remote sensing situation where the UAVs send their individual information to a UAV node designated for data fusion and overall management of the network. 
We assume that the UAV network does not own spectrum and opportunistically uses an external spectrum to complete the remote sensing mission. We present an algorithm based on team reinforcement learning to maximize the total utility of the system by re-positioning and determining the proper task for each drone, while minimizing the number of steps to achieve the optimal configuration. We provide 
experimental results to verify the convergence of the proposed algorithm.






\section{System Model}\label{System_Model}
Let us assume a network of $N$ UAVs in a proximity of a licensed transmitter-receiver pair that are willing to share their spectrum with the UAV network in exchange for cooperative relaying service. The primary network is considered as a terrestrial network; while the secondary network consists of the UAVs that hover in a constant altitude. The primary pair may be in need of cooperative relaying services for experiencing a low quality direct transmission due to several reasons such as long distance between the transmitter and receiver, shadowing and fading effects. In this paper, we assume that there is no direct transmission link between the PU's transmitter and receiver, however, the proposal spectrum sharing method can be used in scenarios where the PU participates in spectrum leasing to take advantage of cooperative diversity gain.
We assume that the UAVs are managed by a high altitude platform (HAP) UAV called as \textit{emergency center} located in a fixed location \cite{mousavi2018leader}. 

The UAV network owns a dedicated frequency bandwidth for control commands and exchanging critical information, however it requires additional spectrum to transmit high resolution video and images during a disaster monitoring mission. In order to provide additional on-demand spectrum, the UAVs can participate in a cooperative spectrum leasing transaction with the primary network. In this case, the PU will lease half of its time access to its spectrum to the sensing UAVs, and the relay UAVs will assist the PU with forwarding their packets during the other half of the time slot. In other words, the UAVs can serve in two \textit{sensing} and \textit{relaying} modes. Here, we consider a centralized control scenario in which the emergency center determines the action of each UAV in terms of their motion and their role (relaying or sensing) to maximize the network utility.

Let us consider a case where $K$ UAVs perform as relay, and the remaining $N - K$ perform the sensing task. An example of this scenario with $N=6$ and $K=3$ is depicted in Fig. \ref{fig:system_grid_1}. 
In this model, we consider a 2-D random walk for the UAVs, so that they can change their locations into adjacent cells in the grid using one of the four movement actions $\{ \textnormal{Up}, \textnormal{Down}, \textnormal{Left}, \textnormal{Right} \}$. The UAVs are not allowed to take diagonal steps.

The channels between all nodes including the UAVs, the emergency center, and the PU are calculated based on pairwise distances during each time slot.
Following similar works including 
\cite{afghah2018reputation}, we assume that CSI parameters are known for the HAP UAV. Channel coefficient for each pair of users is defined as follows: i) $h_{PT, U_i}$ is the channel parameter between the primary transmitter and $i^{th}$ UAV, ii) $h_{U_i, PR}$ carries the channel information between the $i^{th}$ UAV and the primary receiver, and iii) $h_{S, U_i}$ and $h_{U_i, E}$ denote the channel parameters between the $i^{th}$ UAV and the source and the emergency center respectively.
Moreover, we assume that the noise terms at the receiver sides are normally distributed symmetric complex values, $Z \sim CN(0, \sigma^2)$. 

In the literature, many papers including \cite{shamsoshoara2015enhanced, mohammadi2018qoe}
address power optimization; while we assume that the transmission power is arbitrary but constant for all transmitters.

\begin{figure}[bht]
\centering
\includegraphics[width=0.3\linewidth]{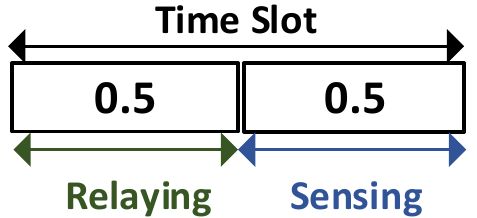}
\caption{One time-slot for both networks}
\label{fig:timeslot}
\end{figure}

Based on Fig. \ref{fig:timeslot}, we divide each time slot into two parts. We assume that, relaying UAVs receive the primary packet and forward them to the primary receiver in the first half. And in the second half, sensing UAVs forward the collected data to the emergency center. In each half slot, UAVs utilize full-duplex for the transmission. 

The received signal for the UAV can be written as follow: 
\begin{align}\label{eq:Y_U}
y_{u}[n] = h_{i, u}x_{i}[n] + z_{r}[n]
\end{align}
In (\ref{eq:Y_U}), subscript $i$ can be source or PU transmitter, $x_i$ is the signal sent by by the PU transmitter or the source and $y_u$ is the received signal at UAV $u$. On the other side of the transmission, the received signal for the emergency center or the primary receiver is written as follow: 
\begin{align}\label{eq:Y_dest}
y_{j}[n] = h_{u, j}x_{u}[n] + z_{j}[n],	
\end{align}
where subscript $j$ can represent the emergency center or the PU-receiver, $x_u$ is the transmitted signal by UAV $u$, and $y_j$ is the received signal at the $j^{th}$ receiver which can be the emergency center or the PU-receiver. In both (\ref{eq:Y_U}) and (\ref{eq:Y_dest}), $h_{ij}$ defines the effect of path loss between nodes $i$ and $j$. 



In this work, we utilized the Amplify and Forward (AF) relaying mode at the relay UAVs. 
Based on \cite{yu2005amplify, nosratinia2004cooperative}, the throughput for the emergency center receiver can be obtained as:
\begin{align}\label{eq:R_SE}
R_{SE} & = \frac{1}{2}\log_2(1 + P_{S}|h_{S,E}|^2  \\
\nonumber
&+\frac{P_{S}|h_{S,U_i}|^2\: P_{U_i}|h_{U_i, E}|^2}{\sigma^2 +P_{S}|h_{S,U_i}|^2 + P_{U_i}|h_{U_i, E}|^2}),
\end{align}
where $P_S$ and $P_{U_i}$ are transmission power for source and UAV respectively.
$\sigma^2$ is the background noise power. $i$ denotes the index for the relay(UAV). On the other hand, the throughput rate for the primary receiver is shown as:
\begin{align}\label{eq:R_PU}
R_{PU} & = \frac{1}{2}\log_2(1 + P_{PT}|h_{PT,PR}|^2  \\
\nonumber
&+ \frac{P_{PT}|h_{PT,U_i}|^2 \: P_{U_i}|h_{U_i, PR}|^2}{\sigma^2 +P_{PT}|h_{PT,U_i}|^2 + P_{U_i}|h_{U_i, PR}|^2}),
\end{align}
where $P_{PT}$ is the primary transmitter power.
It is noteworthy that (\ref{eq:R_SE}) and (\ref{eq:R_PU}) are true for single relay scenarios.
This system involves a multi-agent scenario including $K$ relay UAVs and a set of $N-K$ sensing UAVs, hence the transmission rate of the relaying and sensing sub-systems can be defined as follows if $U_E = \{E_1, E_2, \dots E_{N-K} \}$ denotes the sensing set and $U_R = \{R_1, R_2, \dots, R_{K}\}$ is the set of relaying UAVs:

\begin{align}\label{eq:R_SE_multi}
R_{SE} & \textnormal{(Multi-UAV)}  = \frac{1}{2}\log_2(1 + P_{S}|h_{S,E}|^2  \\
\nonumber
&+ \sum_{j \in U_E}^{} \frac{P_{S}|h_{S,U_j}|^2 \: P_{U_j}|h_{U_j, E}|^2}{\sigma^2 +P_{S}|h_{S,U_j}|^2 + P_{U_j}|h_{U_j, E}|^2}),
\end{align}
\begin{align}\label{eq:R_PU_multi}
R_{PU} & \textnormal{(Multi-UAV)} = \frac{1}{2}\log_2(1 + P_{PT}|h_{PT,PR}|^2  \\
\nonumber
&+ \sum_{l \in U_R}^{} \frac{P_{PT}|h_{PT,U_l}|^2 \: P_{U_l}|h_{U_l, PR}|^2}{\sigma^2 +P_{PT}|h_{PT,U_l}|^2 + P_{U_l}|h_{U_l, PR}|^2}),
\end{align}
where, $R_{SE}(\textnormal{Multi-UAV})$ and $R_{PU}(\textnormal{Multi-UAV})$ are the achievable rates using multi UAVs. Since we assumed the distances between the source and emergency center and also PT and PR are long, we can eliminate both $h_{S,E}$ and $h_{PT, PR}$ terms from the rate ($|h_{S,E}| \simeq 0 \textnormal{ and } |h_{PT, PR}| \simeq 0 $).
At the end of each time slot, the PU receiver and the emergency center announce their received throughput to the UAVs in order to provide them with a feedback (i.e., reward) based on the previous states and their taken actions.

Based on the definitions in our system model, we can define the reward or gain as a throughput difference for the whole system between to consecutive time slots such as $(t)$ and $(t - 1)$:
\begin{align}\label{eq:reward}
Reward & =\beta_{1} \times (R_{PU}(t) - R_{PU}(t-1)) \\
\nonumber
& +\beta_{2} \times(R_{SE}(t) - R_{SE}(t-1)),
\end{align}
where, $\beta_{1}$ and $\beta_{2}$ are two control parameters. Reward in (\ref{eq:reward}) is obtained based on the location of the UAVs, their tasks, and the distance to their destinations or from their sources. The goal of our proposed model is to find the best location for UAVs and the right task that result in an optimal summation throughput rate. In Section \ref{Learning}, we use this reward function to develop a reinforcement technique for task allocation and motion planning.
\section{Multi-Agent Team Reinforcement Learning}\label{Learning}
Multi-Agent Reinforcement Learning (MARL) algorithms can be categorized into three branches based on the nature of the problem. First, all the agents work cooperatively with each other. In this branch, all the agents have the same goal of maximizing the common discount factor. Also, they all have the same reward function. In the second category, the agents are fully competitive which is known as zero-sum games. In third one, the behavior of agents is a mixture of two previous groups which is neither competitive nor cooperative. Our problem in this paper is the type with fully cooperative agents. Two approaches to solve these kinds of problems are Nash Q-learning and team learning \cite{littman2001friend, choi2010survey}.
We aim to use the team Q-learning for our UAVs. In the team learning approach, a single learner such as the emergency center monitors the behavior of all agents. The reason we chose this algorithm is that the single learner can utilize simple machine learning algorithms such as Q-learning in the multi-agent learning environment similar to the scenario in this paper.
To model a single UAV agent in the learning environment, the Markov Decision Process (MDP) is investigated. In each step, the UAV interacts with its environment which is the state of the problem. In this specific problem, the state for the UAV is the location on the grid size.
Let $S_t \in \mathcal{S}$ denote the state for each UAV at step $t$, where $\mathcal{S}$ is the set for all possible states. Time step $t$ is different
and it depends on the grid size of the UAVs. The $i^{th}$ UAV at each state such as $s_t^{i}$ can follow an action $a_t^{i}$ based on a policy form an action set $\mathcal{A}$. The action set $\mathcal{A}$ is $\{$$\textnormal{Up(0)}$, $ \textnormal{Down(1)}$, $\textnormal{Left(2)}$, $\textnormal{Right(3)}$, $\textnormal{Emergency Task(4)$, $ \textnormal{Relaying Task(5)}}$$\}$. The first four actions change the location for the UAV and keep the previous task from the previous location. However, the last two actions, keep the UAV in same location, but decide on its task. Action $a_t^{i}$ changes $S_t$ to $S_{t+1} \in \mathcal{S}$ for the whole system. Based on the new state $S_{t+1}$, the updated reward $r_t \in \mathcal{R}$ is awarded to the system based on the (\ref{eq:reward}). In a single-UAV environment, the goal of the UAV is to find the optimal policy which chooses actions that maximize the $E(\sum_{t=0}^{\textnormal{Number of Steps}}\gamma^t r_t)$ standing for expected discounted reward over time where $\gamma$ is a discount factor. SARSA \cite{rahili2018optimal} and Q-learning \cite{mousavi2017applying}
are two general algorithms for the RL problems. Each MPD consists of a 4-tuple $(\mathcal{S}, \mathcal{A}, {P_a}, {R_a})$ where, $\mathcal{S}$ is a finite set of states, $\mathcal{A}$ is a finite set of actions, $P_a: S \times a \times S' \rightarrow [0, 1]$ or $P_a(S, S') = Pr(S_{t+1}=s' | S_t = s, a_t = a)$ is the probability that choosing an action $a$ for a single UAV will change its state from $S$ to $S'$ at the time $t+1$. $R_a: S \times a \rightarrow \mathbb{R}$ or $R_a(S, S')$ is the expected awarded reward when a single UAV changes its state from $S$ to $S'$ due to action $a$. This immediate reward is obtained based on (\ref{eq:reward}) in this special problem. Then, the single agent can update its Q-value based on:
\begin{align}\label{eq:Qupdate}
Q(S, a) = & (1-\alpha)\times Q(S, a)  \\
\nonumber
+ & \alpha [R_a(S, S') + \gamma \max_{a' \in \mathcal{A}} Q(S', a')],
\end{align}
where $\alpha (0 < \alpha \leq 1)$ is the learning rate. These equations and relations are valid for a single-learner. However, in multi-agent environment, multiple UAVs follow  the taken action at the same time without sharing any information with each other. Hence, the action vector such as $A_t = (a_1, a_2, \dots, a_N)_{t+1} \in \mathcal{A}$ leads the system from $S_t$ to $S_{t+1}$ for all UAVs. Using team Q-learning, (\ref{eq:Qupdate}) can be modified and written as:
\begin{align}\label{eq:TeamUpdate}
Q(S, [a_1, & a_2, \dots, a_N]) =  (1-\alpha)\\
\nonumber
\times & Q(S, [a_1, a_2, \dots, a_N])  \\
\nonumber
+ & \alpha [R_a(S, S') + \gamma \max_{A' \in \mathcal{A}} Q(S', A')],
\end{align}
where $A' = [a_1', a_2', \dots, a_N']$ is the action vector when the UAVs are placed in state $S'$ and $Q$ is the action-value for the whole system.
There are three kinds of rewards based on the (\ref{eq:reward}). If the throughput summation in a new state is greater than the previous state, then the positive reward is given to the system. If the difference is zero, then there is no reward for that (zero value). If the difference is negative between the new state and the previous state, then a negative reward is obtained by the system. 
The issue for team Q-learning is the process of exploration. Considering a 10$\times$10 grid size for 2 UAVs defines 100 states for each UAVs; then, defining a team of 2 UAVs makes 10,000 states for the system just for locations without considering any tasks. This huge number of states can be time-consuming for the learning process.
Action selection method is another important module for the Q-learning which stands for selecting the actions which the UAVs will perform to proceed the learning process. These methods base a fulcrum for exploration and exploitation searches \cite{sutton2018reinforcement}. In this paper, we used $\epsilon$-greedy exploration with a constant value for the $\epsilon$ which means the system chooses random actions with the probability of $\epsilon$ and selects the best actions based on the maximum value in the Q-table with the probability of $1 - \epsilon$. This method is defined in:
\begin{align}\label{eq:epsgreedy}
& [a_1, a_2, \dots, a_N]_{t+1}: \\
\nonumber
&  A_{t+1} = 
\begin{cases}
argmax_A Q(S, A), \qquad \  \textnormal{Probability } 1- \epsilon, 
\\
\textnormal{Random Actions,} \qquad \ \ \ \,  \textnormal{Probability } \epsilon,
\end{cases}
\end{align}
where $N$ is the number of UAVs and $A$ is the action vector for $N$ UAVs which brings the maximum Q value.  
Table \ref{tab:TeamRL} summarizes the mapping information for the team Q-learning and the definitions in this specific problem.
\begin{table}[htbp]
\caption{Mapping the components between the team Q-learning and the system model for a 6$\times$6 grid.}
 \centering{
\label{tab:TeamRL}
\resizebox{0.80\linewidth}{!}{  
\begin{tabular}{c|c|c}
\toprule
\toprule
\Topspace
\Bottomspace
\textbf{\Large Component} & \textbf{\Large Description} & \textbf{\Large Sample size} \\
\hline
\Topspace
\Bottomspace
\textbf{\fontsize{14}{17}\selectfont Actions} & \fontsize{14}{17}\selectfont Making decisions about moving or the task & \fontsize{14}{17}\selectfont 6
\\
\hline
\Topspace
\Bottomspace
\textbf{\fontsize{14}{17}\selectfont \# of Agents} & \fontsize{14}{17}\selectfont Number of UAVs & \fontsize{14}{17}\selectfont 2
\\
\hline
\Topspace
\Bottomspace
\textbf{\fontsize{14}{17}\selectfont State} & \fontsize{14}{17}\selectfont Considering a grid for all agents together & \fontsize{14}{17}\selectfont $(36)^2=1,296$
\\
\hline
\Topspace
\Bottomspace
\textbf{\fontsize{14}{17}\selectfont Q-Table} & \fontsize{14}{17}\selectfont State(Location), Action(Movement or Task) values & \fontsize{11}{13}\selectfont $(36 \times 6)^2 = 46,656$
\\
\hline
\Topspace
\Bottomspace
\textbf{\fontsize{14}{17}\selectfont Reward} & \fontsize{14}{17}\selectfont Difference of throughput & \fontsize{14}{17}\selectfont $R(S,A,S') \rightarrow \mathbb{R}$
\\
\bottomrule 
\end{tabular}
} 
} 
\end{table}
In the past few years, deep learning has absorbed lots of attention in different areas such as biomedical signal processing \cite{mousavi2018inter,mousavi2019sleepeegnet,mousavi2018ecgnet}, pattern recognition, computer vision, autonomous vehicles, and natural language processing. Moreover, in recent studies, it has been shown that deep learning tools can be utilized alongside with reinforcement learning challenges to give a bright consciousness of the environment to the agents \cite{mousavi2017traffic,mousavi2016deep,mousavi2016learning,mousavi2017applying,mousavi2018thesis}.   
\section{Simulation Results}\label{Simulation}

In the following simulations, we consider a ground-based PU transmitter-receiver pair as well as a pair of drone-based source and emergency center with randomly initialized locations. We split the tiled coverage area into  $L_1\times L_2$ tiles, where drones are allowed to move in four directions $\{ \textnormal{Up}, \textnormal{Down}, \textnormal{Left}, \textnormal{Right} \}$. Channels between nodes are calculated based on the $h_{j,j} \sim CN(0,d^{-2}_{i,j})$, where $d_{i,j}$ is Euclidean distance between nodes $i$ and $j$. We arbitrarily use $\beta_1=1$, $\beta_2=1$ (defined in Eq. \ref{eq:reward}), $\alpha=0.1$, $\gamma=0.3$, $\epsilon=0.1$, $\sigma^2=1nW$, $P_{PT}=10mW$, $P_S=20mW$, and $P_{U_i}$=$20mW$, unless otherwise specified explicitly. In the $\epsilon$-greedy exploration, we consider a constant rate for the exploration and exploitation phases.
The number of states is $(L_1\times L_2)^{N}$ for $N$ drones. 
We execute the MARL algorithm for 200 episodes to fill in the Q-tables for all drones based on their actions and locations. 
We use sequential episodes, where the resulting Q-tables for each episode is used to initialize Q-tables for the next episode.  
The number of steps per episode to cover the entire state space is calculated based on the experience and the number of states. We then repeat each experiment in 10 external iterations for enhanced accuracy and avoiding bias to initialization. Timing value for each scenario is provided for Python implementation with Intel Xeon(R) CPU @ 3.50GHz and 16.0GB RAM. 
The simulation parameters are summarized in Table \ref{tab:sim}.


\begin{table*}[htbp]
\caption{Simulation parameters and required times for the simulation with 2 UAVs and 6 actions.}
 \centering{
\label{tab:sim}
    \resizebox{1\linewidth}{!}{  
\begin{tabular}{c|ccccccc}
\toprule
\toprule
\textbf{Grid Size:} & {2$\times$2 (4)} & {3$\times$3 (9)} & 4$\times$4(16) & {5$\times$5 (25)} & {6$\times$6 (36)} & 8$\times$8 (64) & 10$\times$10 (100) \\
\hline
\textbf{Number of States:} & 4$\times$4=16	& 9$\times$9=81	& 16$\times$16=256 & 625 & 1,296	& 4,096	& 10,000
\\
\hline
\textbf{Size of the Q-Table:} & 576	& 2,916	& 9,216	& 22,500	& 46,656	& 147,456	& 360,000  \\
\hline
\textbf{Number of Iterations:} & 10 & 10 & 10 & 10 & 10 & 10 & 10 \\
\hline
\textbf{Number of Episodes in each Iteration:} & 200	& 200	& 200	& 200 &	200 & 200 &	200 \\
\hline
\textbf{Number of Steps in
 each Episode:} & 120	& 800	& 2,000	& 5,000	& 12,000	& 20,000	& 30,000 \\ 
\hline
\textbf{Process Time for
 each epoch (Sec):} & {0.102628}	& {0.762713}	& {2.425558 } & {8.757846}	& {49.007077}	& {146.936207}	& {291.176670} \\
\hline
\textbf{Total required time:} & {205.256s}	& {25.4m}	& {80.85m}	& {4.8h}	& {27.2h}	& {3 days}	& {6 days}\\
\bottomrule 
\end{tabular}
}
}
\end{table*}
\begin{figure}[hbt]
	\centering
	\includegraphics[width=0.8\columnwidth]{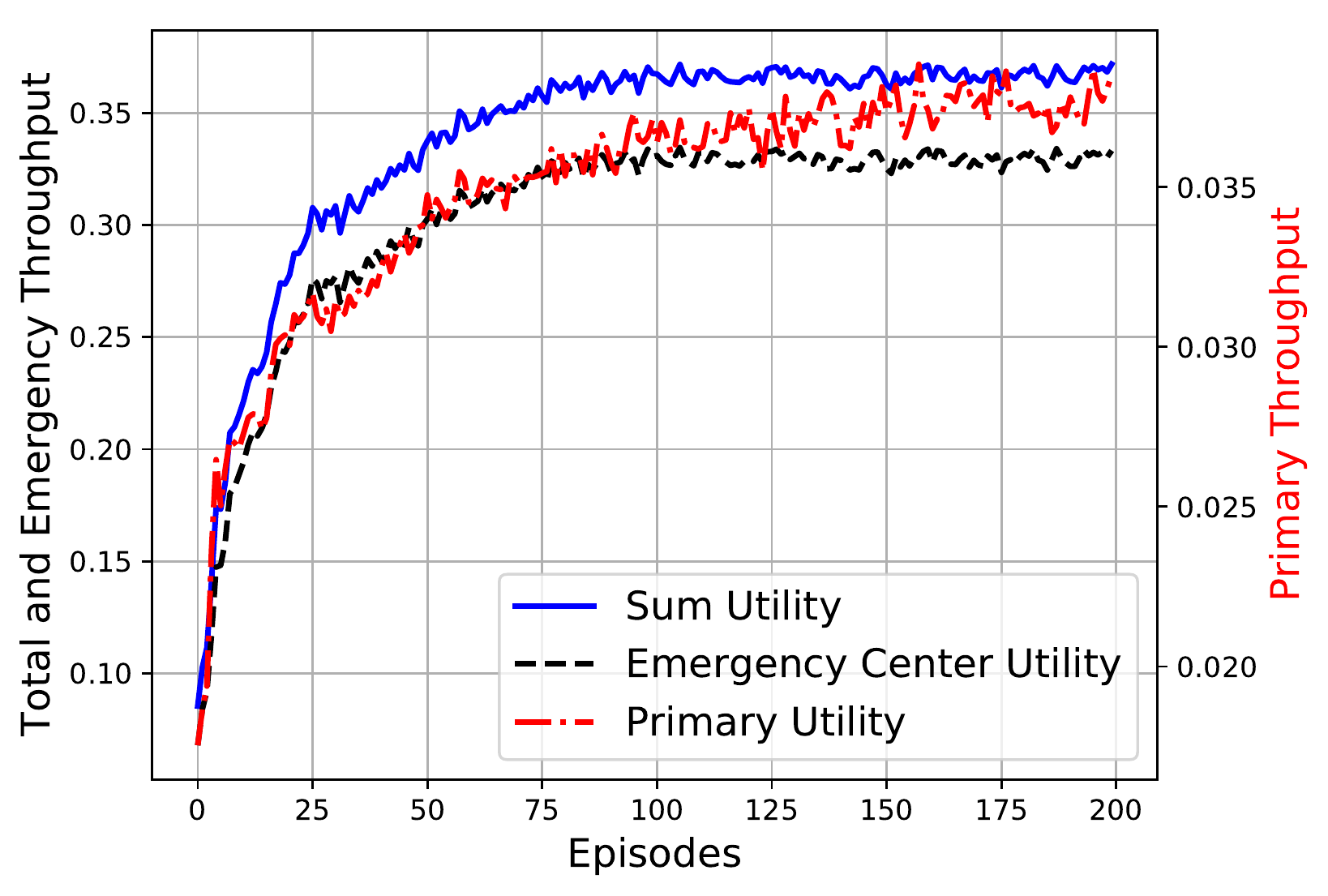}
	\caption{Summation throughput for emergency, primary, and both networks for all UAVs.}
    \label{fig:sumutility}
\end{figure}
Figure \ref{fig:sumutility} shows the sum utility for primary and emergency network and the summations of those for each episode. The sum utility represents the accumulated utility during all steps of the episode. We observe that due to using sequential learning, the latter episodes yield a higher sum utility suggesting that the algorithm converges to its final value faster. This behavior is valid for both primary and secondary networks. 

\begin{figure}[hbt]
	\centering
	\includegraphics[width=0.82\columnwidth]{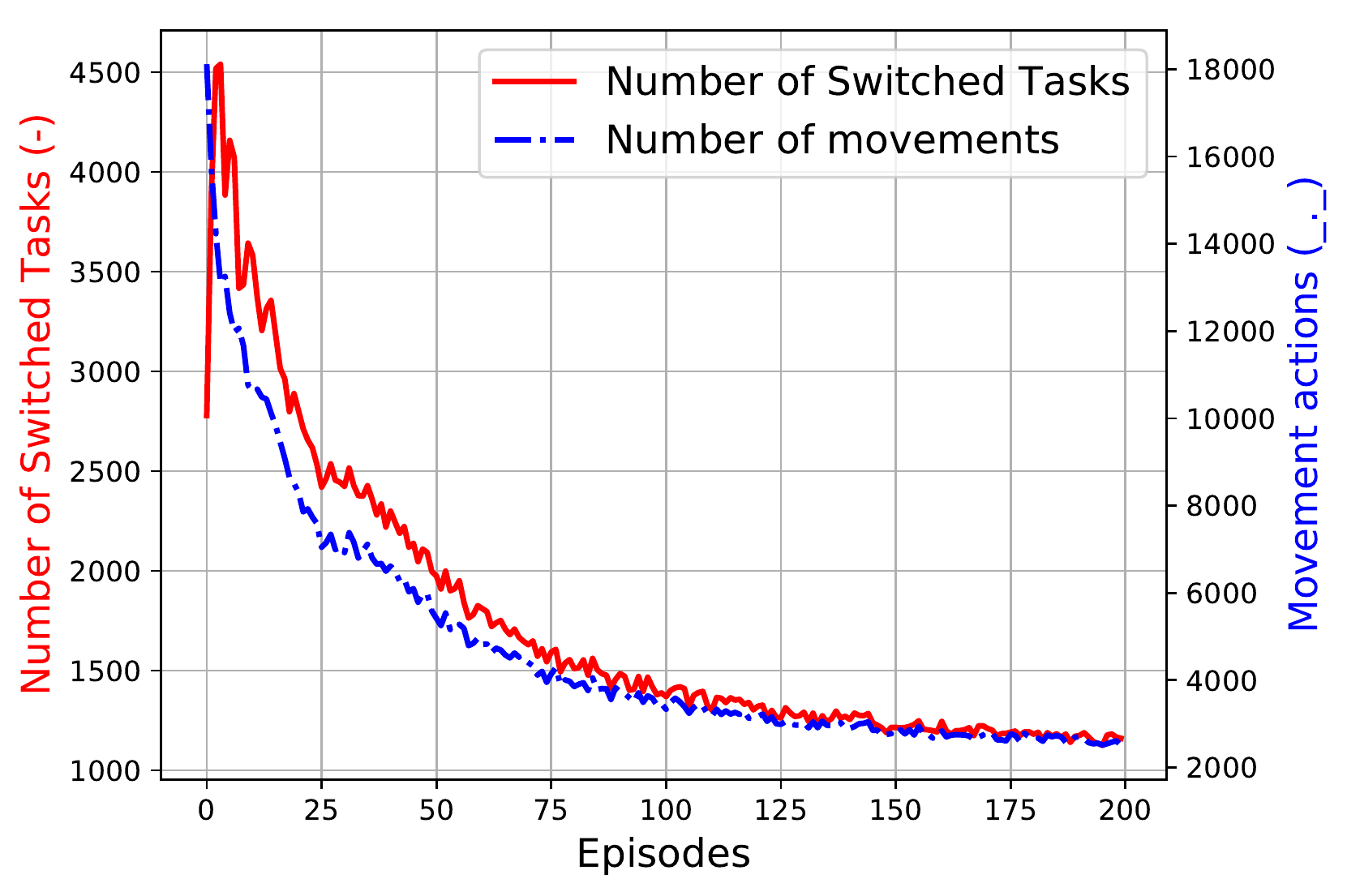}
	\caption{Number of switched tasks and movements for all UAVs in each episode.}
    \label{fig:task}
\end{figure}

\begin{figure}[hbt]
	\centering
	\includegraphics[width=0.8\columnwidth]{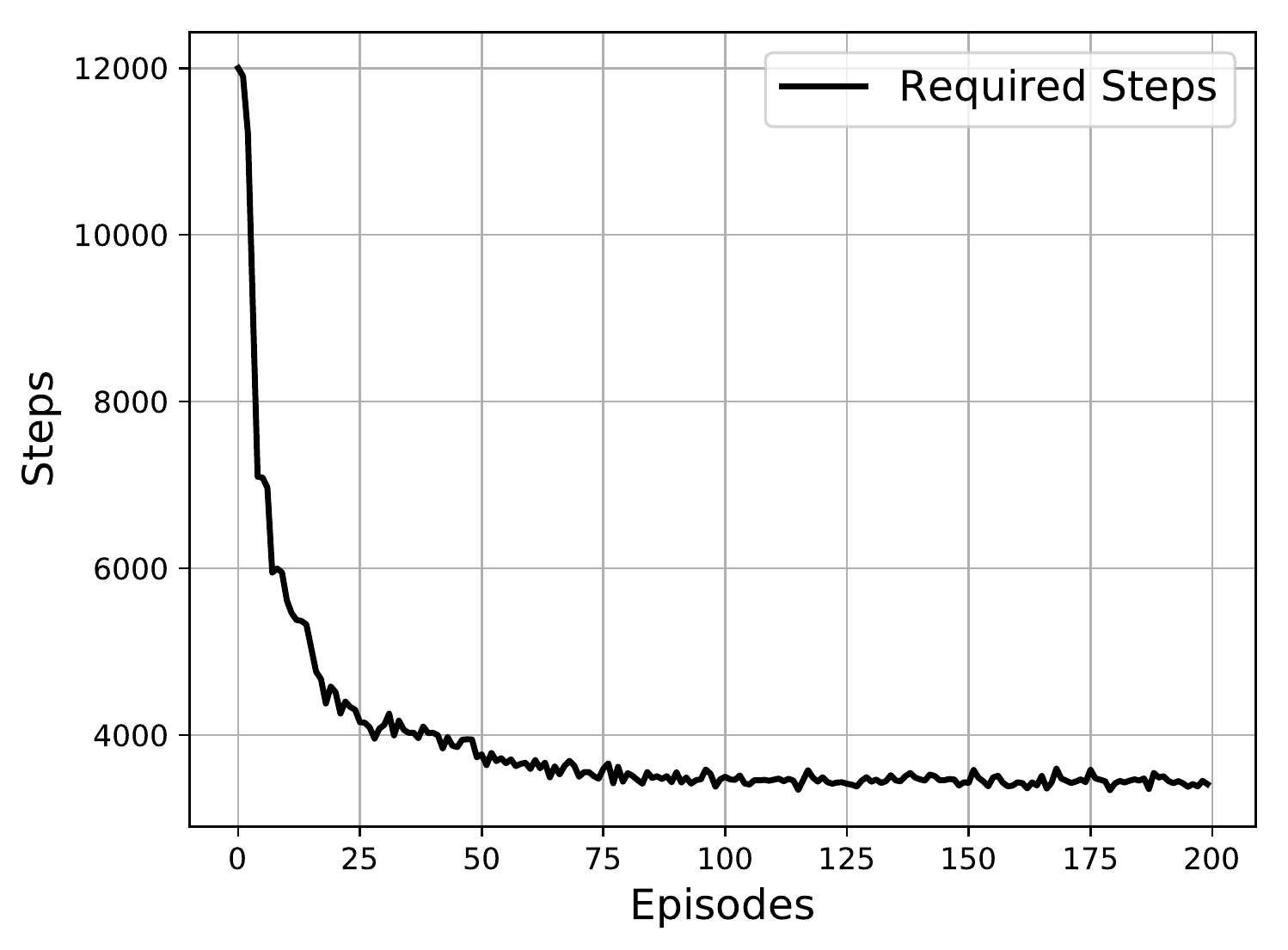}
	\caption{Number of required steps in each episode to get 90\% of the throughput summation.}
	\vspace{-10pt}
    \label{fig:steps}
\end{figure}

Figure \ref{fig:task} demonstrates the number of times that UAVs switched their task and also the number of times they changed their locations in each episode. Each value at each point for a specific episode is the summation for both UAVs for all taken steps. It is logical that the HAP starts searching all states by switching tasks in each location to get a new reward. However, as simulation goes on, the HAP starts learning from its past experiences and switches the tasks for the UAV networks fewer compared to the beginning of each iteration. Also, HAP forces the UAVs to explore more locations in smaller episodes.
For larger episodes, the number for movements or changing location is less than the beginning of the iteration. These behaviors demonstrate the learning process toward the optimal location and task to gain a higher throughput rate for both networks.
Fig. \ref{fig:steps} shows the required steps to get 90\% of the throughput summation in each episode. The number of steps is decreasing when the number of episodes is increasing which explains the learning behavior and denotes that in less necessary steps, the emergency center can find the best states (Location) and actions for all UAVs to get the optimal reward. 
All figures in this section are sketched for the grid 6$\times$6 (36).
Based on Table \ref{tab:sim}, the process time for each iteration is increasing exponentially for larger grid size. Moreover, this process time is dependent to the number of UAVs. This happens due to the nature of the team reinforcement learning. Since we are using distance-based deterministic channel coefficients, offline learning methods can be used to fill in the Q-tables.

\section{Conclusion}\label{Conclusion}
In this paper, we introduced a solution for dynamic spectrum sharing in which a network of search-and-rescue UAVs needs an additional spectrum to send critical information such as videos to the emergency center. In this method, some of the UAVs provide a relaying service for a primary network in exchange for borrowing an additional required spectrum. We proposed a team-learning MARL algorithm, in which the emergency center of the UAV networks can determine the optimal actions of the UAVs in terms of their motion as well as the optimum task. It is worth mentioning that, in practice, the UAV network is trained by this learning mechanism in an offline manner to identify the best set of actions for different situations and use this knowledge for fast decision making during the missions. Simulation results showed the proposed distributed method converges to the optimal locations and tasks for both the primary and emergency network. The optimal state is the win-win solution for both groups. For the future works, we are working on other learning methods to reduce the effect of states (locations and tasks) and agents (UAVs) on the performance of the algorithm. Due to the exponential behavior of team learning, it takes time for offline learning; however, we aim to increase the number of UAVs and the grid size at the same time. 



\bibliography{main}
\bibliographystyle{IEEEtran}

\end{document}